# The Topology of Hardship: Empirical Curriculum Graphs and Structural Bottlenecks in Engineering Degrees

Hugo Roger Paz
PhD Professor and Researcher Faculty of Exact Sciences and Technology National University of Tucumán
Email: hpaz@herrera.unt.edu.ar
ORCID: https://orcid.org/0000-0003-1237-7983

**ABSTRACT**

Engineering degrees are often perceived as "hard", yet this hardness is usually discussed in terms of content difficulty or student weaknesses rather than as a structural property of the curriculum itself. Recent work on course-prerequisite networks and curriculum graphs has shown that study plans can be modelled as complex networks with identifiable hubs and bottlenecks, but most studies rely on official syllabi rather than on how students actually progress through the system (Simon de Blas et al., 2021; Stavrinides & Zuev, 2023; Yang et al., 2024; Wang et al., 2025).

This paper introduces the notion of **topology of hardship**: a quantitative description of curriculum complexity derived from empirical student trajectories in long-cycle engineering programmes. Building on the CAPIRE framework for multilevel trajectory modelling (Paz, 2025a, 2025b), we reconstruct degree–curriculum graphs from enrolment and completion data for 29 engineering curricula across several cohorts. For each graph we compute structural metrics (e.g., density, longest path, bottleneck centrality) and empirical hardship measures capturing blocking probability and time-to-progress. These are combined into a composite hardship index, which is then related to observed dropout rates and time to degree.

Our findings show that curriculum hardness is not a vague perception but a measurable topological property: a small number of structurally dense, bottleneck-heavy curricula account for a disproportionate share of dropout and temporal desynchronisation. We discuss implications for curriculum reform, accreditation, and data-informed policy design.

# 1. INTRODUCTION

Engineering programmes are widely regarded as demanding, selective, and structurally unforgiving. Policy debates around attrition usually invoke individual preparedness, motivation, or socio-economic vulnerability, while the curriculum itself is treated as a neutral backdrop. However, a growing body of work suggests that the *structure* of the curriculum—its sequencing, prerequisite chains, and concentration of key courses—plays a central role in shaping student pathways, time to degree, and dropout patterns (Simon de Blas et al., 2021; Yang et al., 2024; Wang et al., 2025).

## 1.1 Curriculum as a complex network

Curriculum analytics has increasingly adopted tools from network science to represent programmes as graphs where nodes are courses and edges capture prerequisite or recommended-before relationships (Stavrinides & Zuev, 2023; Yang et al., 2024). These course-prerequisite networks reveal hidden structures such as hubs, bridges, and bottlenecks, and can be used to support curriculum design, resource allocation, and student navigation. Recent studies have used network metrics to characterise curricular complexity, identify influential courses, and compare structural patterns across institutions and disciplines (Alrizqi & Godwin., 2025; Wang et al., 2025).

Beyond static catalogues, curriculum graphs have been linked to outcome measures. For example, network-based models have been used to support course sequencing decisions and redesign efforts in outcome-based education (Simon de Blas et al., 2021; Nazyrova et al., 2023). Other work has embedded curriculum representations into knowledge graphs and recommendation systems for aligning programmes with employer demands and skills frameworks (Aljohani et al., 2022; Li et al., 2024). These strands converge on a shared view: curricula are not simple lists of modules, but interconnected systems whose topology matters.

## 1.2 From official plans to empirical curriculum graphs

Most existing curriculum graph studies rely on *official* prerequisite information extracted from syllabi, handbooks, or accreditation documents (Stavrinides & Zuev, 2023; Yang et al., 2024). This approach is appropriate when formal prerequisite structures are well specified and faithfully implemented. However, in many public universities—and particularly in long-cycle engineering programmes—students' trajectories often diverge substantially from the theoretical design. Local enforcement of prerequisites, informal advice, resource constraints, and regulatory exceptions can generate a de facto curriculum that differs markedly from the declared plan.

Recent work has begun to address this gap by constructing curriculum graphs directly from student flow data rather than from catalogues. Baucks et al. (2022) proposed deriving edges from observed enrolment sequences, enabling simulation of policy changes even in prerequisite-free curricula. Similarly, broader curriculum analytics approaches emphasise the need to compare "process maps" of actual student behaviour with the intended structure of the programme (Qushem et al., 2025). Yet, empirical graphs built from trajectories are still rare, and even less work has systematically related their topological properties to dropout and time-to-degree at scale.

**1.3 Curriculum hardness as a structural property**

In parallel with graph-based curriculum modelling, learning analytics and educational data mining have produced a rich literature on student dropout prediction and risk profiling. However, most predictive models treat curriculum features as background variables or rely on opportunistic indicators such as number of failed courses or cumulative grade point average, often with substantial risk of data leakage (Paz, 2025a). These models can achieve high accuracy yet offer limited explanatory power for structural reform, because they conflate student characteristics, teaching practices, and curricular design.

Within the CAPIRE framework for multilevel trajectory modelling, curriculum structure is elevated to a first-class component of the analytical architecture (Paz, 2025b). Here, the N2 layer explicitly encodes the topology of the degree plan, while other layers capture entry attributes, performance trajectories, and macro-context. This perspective motivates a shift from talking informally about "hard" or "blocked" programmes to measuring **curriculum hardness** as a property of the degree–curriculum graph itself. Intuitively, a curriculum may be considered structurally "hard" if it presents long critical chains, high dependency density, or a concentration of high-risk bottleneck courses.

The central idea of this paper is to formalise this intuition as the **topology of hardship**: a composite characterisation of curriculum complexity that combines structural graph metrics with empirical measures of blockage and progression derived from student trajectories.

**1.4 Research gap and contribution**

Despite promising developments in course-prerequisite networks and data-driven curriculum design, three limitations remain:

1. **Dependence on official prerequisites.** Most studies assume that the declared prerequisite structure accurately describes how students move through the programme, which is often unrealistic in open or flexible systems (Yang et al., 2024; Baucks et al., 2022).

2. **Limited linkage to longitudinal outcomes.** While some work uses network metrics to identify "important" courses, few analyses systematically relate curriculum topology to multi-year outcomes such as dropout and time-to-degree in engineering programmes (Alrizqi & Godwin, 2025).

3. **Insufficient attention to degree–curriculum granularity.** Many analyses aggregate across degrees or curriculum reforms, masking differences between generations of study plans and between disciplines inside engineering faculties.

This study addresses these gaps by:

- reconstructing **empirical curriculum graphs** for 29 engineering degree–curriculum combinations using longitudinal student data from a large public university;

- defining a set of **structural** and **empirical hardship** metrics that quantify blocking probability, bottleneck centrality, and critical chain length at degree–curriculum level; and

- relating a composite **hardship index** to observed dropout rates and time-to-degree across programmes.

By grounding curriculum graphs in observed trajectories rather than in official plans, and by connecting their topology to institutional outcomes, the paper contributes a methodology for treating curriculum hardness as a measurable, actionable property of the educational system. This, in turn, supports data-informed curriculum redesign, accreditation processes, and scenario analysis within broader frameworks such as CAPIRE.

## 2. DATA AND INSTITUTIONAL CONTEXT

### 2.1 Institutional setting

The study is based on longitudinal registry data from a large public university of engineering and applied sciences in Latin America. The institution offers long-cycle engineering degrees (five or more years) in areas such as Civil, Mechanical, Electrical, Electronics, Industrial and Computer Engineering. Entry is open and tuition-free, but progression is strongly shaped by internal regulations on regularity, examination regimes and promotion rules.

The data infrastructure is built around an institutional academic records database, which stores complete histories for all students: degree enrolments, curriculum versions, course registrations, examination results, equivalences and graduation events. For the purposes of this paper, we treat the degree as the primary programme entity and the **curriculum** (or "plan") as the specific version of the study plan that was in force at the time of a student's enrolment or transfer.

**2.2 Data sources and preprocessing**

All analyses use the CAPIRE multilevel data layer already developed for earlier work on student trajectories and retention. This layer draws data from the SQL Server instance that hosts the institutional database and exports it into a set of normalised tables in parquet format, with explicit handling of temporal information, episode boundaries and leakage control. The N2 institutional layer, which underpins the present study, operates at the level of degree, curriculum and course.

From the raw registry we extract, for each student:

- degree identifier;

- curriculum (plan) identifier;

- course code;

- type of academic event (course registration, regular exam, free exam, promotion, equivalence);

- result (pass, fail, absent, withdrawal);

- grade where available;

- academic period (year and, when applicable, semester);

- status changes such as degree transfer, curriculum change and graduation.

These records are then transformed into **spells**, defined as maximal time segments where a student remains in the same (degree, curriculum) combination. Whenever a student changes degree or moves to a new curriculum within the same degree, a new spell begins. This segmentation ensures that trajectories are analysed relative to the correct degree–curriculum unit and prevents contamination between plans.

For each spell we derive a **subject_attempts** table where each row corresponds to a single attempt at a course (including registrations leading to failure or absence) and a **subject_outcomes** table summarising the full history of that student–course pair (first attempt period, pass period, number of attempts, ever-passed flag). These two tables provide the empirical basis for reconstructing curriculum graphs and estimating blocking probabilities.

**2.3 Degree–curriculum units and sample selection**

The empirical analyses operate at the level of degree–curriculum units, defined as the combination of a degree programme and a specific curriculum (or "plan") in force during a given period. The historical registry contains **169** such combinations across engineering and exact sciences, spanning several decades of institutional reforms. However, not all of these units have sufficient student activity or internal connectivity to support reliable reconstruction of empirical curriculum graphs, a challenge commonly reported in curriculum analytics based on administrative records (Heileman et al., 2018; Paz, 2025a).

Several factors motivate a conservative selection strategy. Some early curricula appear in the digital archive with only a handful of enrolled students, typically because systematic electronic recording began after the plan was already being phased out. Other plans were formally approved but never fully implemented, or exhibit inconsistent course codes and idiosyncratic equivalence records. Including such units would produce graphs with very few nodes and edges, whose topological metrics would be dominated by noise rather than by meaningful curricular structure (Baucks et al., 2022; Yang et al., 2025).

To address these issues, we apply two filters before entering any unit into the comparative analysis. **First**, a *subject validity filter* uses the subject catalogues derived from the CAPIRE data layer (Paz, 2025a) to flag anomalous courses. For each course we compute empirical statistics such as number of distinct students, number of attempts and temporal span. Courses with extremely low activity, placeholder-like codes or inconsistent mappings across curricula are treated as invalid and excluded from graph construction. **Second**, a *graph sufficiency filter* is applied to each provisional degree–curriculum graph. Units with fewer than a minimum number of nodes and edges (operationalised here as fewer than eight distinct subjects or fewer than five inferred dependencies) are labelled *degenerate* and removed from subsequent analyses, as their topology cannot be meaningfully interpreted in terms of curriculum hardness (Heileman et al., 2018; Krasnov, 2024).

After applying these filters we retain a **robust analytic sample of 29 degree–curriculum units** with adequate enrolment, non-trivial internal structure and complete outcome information for the cohorts considered. These 29 units cover the main engineering disciplines offered by the institution and multiple generations of curricula within some degrees, enabling both cross-disciplinary comparisons and within-degree contrasts between older and reformed plans. All structural metrics, hardship indices and outcome relationships reported in Sections 3 and 4—and all associated tables and figures—are computed exclusively on this filtered set of 29 degree–curriculum units.

**2.4 Outcome measures**

To relate curriculum topology to institutional performance, we compute two outcome measures for each degree–curriculum:

- **Dropout rate.** Defined as the proportion of students in the entry cohorts considered who exit the degree without graduating and do not reappear in the same degree–curriculum within a specified follow-up window. Students who switch to another degree are treated as dropouts for the source degree but may re-enter the analysis under a new spell in the destination degree.

- **Mean time to degree.** For students who graduate within the observation window, we measure elapsed time in academic years from initial enrolment in the degree to graduation. This is summarised at degree–curriculum level as the arithmetic mean of completed times.

These outcomes are computed from the same spell-based longitudinal representation as the curriculum graphs, ensuring internal consistency. Importantly, hardship metrics are derived from course-level structure and progression patterns, while dropout and time-to-degree are treated as external aggregate outcomes; this separation avoids circularity and supports a genuinely structural interpretation of curriculum hardness.

## 3. ANALYTICAL FRAMEWORK: FROM TRAJECTORIES TO THE TOPOLOGY OF HARDSHIP

This section describes how we move from raw student trajectories to empirical degree–curriculum graphs, and how we derive structural and hardship metrics at both course and programme level. The pipeline follows the CAPIRE curriculum graph architecture (Paz, 2025a) while adopting flow-based techniques for graph inference from student data (Baucks et al., 2022; Yang et al., 2025; Zuev, 2025).

### 3.1 Empirical degree–curriculum graphs

#### 3.1.1 Spells and course histories

Using the spell representation introduced in Section 2, we restrict attention to records where both degree and curriculum remain constant. For each (degree, curriculum) we construct:

- a **subject_attempts** table with one row per student–course–attempt;
- a **subject_outcomes** table summarising, for each student–course pair, first attempt period, pass period (if any), number of attempts and pass indicator.

#### 3.1.2 Empirical course levels

We define an empirical "level" for each course $j$ in a given degree–curriculum as the median of the first-attempt period among all students who ever enrolled in $j$ within that unit. Periods are measured in academic years since initial enrolment in the degree.

$$\text{level}_j = \text{median}(t_{ij}^{\text{first}}),$$

where $t_{ij}^{\text{first}}$ is the academic year in which student $i$ first attempts course $j$. This scalar level provides a coarse ordering of courses that reflects actual sequencing rather than catalogue position, similar in spirit to other empirical progression measures on curricular networks (Yang et al., 2025; Paz, 2025a).

### 3.1.3 Edge inference from student flow

Edges are inferred from observed enrolment and completion sequences rather than from official prerequisite lists, following the idea of constructing curriculum graphs from measurable student flow (Baucks et al., 2022).

For a pair of distinct courses $(i, j)$ within the same degree–curriculum, we consider only students who enrolled in both. We then estimate:

- the probability that $i$ is attempted before $j$:

$$p_{\text{order}}(i \prec j) = P(t_i^{\text{first}} < t_j^{\text{first}}),$$

- the probability of passing $j$ without ever passing $i$:

$$p_{\text{bypass}}(j \mid \neg i) = P(\text{pass } j \wedge \neg \text{pass } i).$$

We create a directed edge $i \to j$ if:

1. **Temporal dominance:** $p_{\text{order}}(i \prec j) \geq \theta_{\text{order}}$ (set to 0.7 in this study).
2. **Limited bypass:** $p_{\text{bypass}}(j \mid \neg i) \leq \theta_{\text{bypass}}$ (set to 0.2).

Intuitively, an edge exists when most students encounter $i$ before $j$, and passing $j$ without having passed $i$ is rare. This captures *effective* rather than formal prerequisites and is compatible with earlier work on empirical curriculum graphs in open or prerequisite-free systems (Baucks et al., 2022).

### 3.1.4 Enforcing acyclicity and robustness

The edge selection rules can still produce cycles due to noise or concurrent enrolment. We therefore construct an initial directed graph and identify strongly connected components. Within each component we remove the smallest number of edges with weakest empirical support (lowest $p_{\text{order}}$, highest $p_{\text{bypass}}$) until the graph becomes a directed acyclic graph (DAG). This follows standard practice in curriculum prerequisite networks and curricular analytics (Stavrinides & Zuev, 2023; Heileman et al., 2018).

To ensure that only informative graphs enter the comparative analysis, we apply the **graph sufficiency filter** described in Section 2: graphs with fewer than eight nodes or fewer than five edges are labelled degenerate and excluded.

**3.2 Structural metrics**

For each non-degenerate empirical curriculum graph we compute a set of structural metrics commonly used in curriculum network analysis (Simon de Blas et al., 2021; Stavrinides & Zuev, 2023; Zuev, 2025).

1. **Size and density.**
    - Number of nodes $N$ (distinct courses).
    - Number of edges $E$.
    - Graph density $D = E/(N(N-1))$.

2. **Depth and critical chain length.**
   We compute the length of the longest directed path from any entry-level node to any sink node. This approximates the minimal number of sequential "levels" a student must traverse under ideal progression and corresponds to the notion of *delay factor* in curricular complexity (Heileman et al., 2018; Krasnov, 2024).

3. **Betweenness centrality and bottleneck concentration.**
   For each node we compute betweenness centrality on the DAG and identify the top decile as bottleneck candidates. We summarise bottleneck concentration as the percentage of courses and credits contained in this top decile, following recent uses of centrality for identifying curricular hubs and blockers (Yang et al., 2025; Paz, 2025a).

4. **Structural hardship score.**
   We normalise selected metrics (density, longest path, bottleneck concentration) to zero mean and unit variance across all degree–curricula and combine them into a Structural Hardship Index:

   $$H_{\text{struct}} = z(\text{density}) + z(\text{longest path}) + z(\text{bottleneck concentration}),$$

where z(·) denotes standardisation. Higher values indicate more complex and tightly constrained structures. For comparative purposes, we additionally compute z-scores of the three hardship indices at degree–curriculum level; these standardised values are reported in Table 3.

### 3.3 Empirical hardship metrics

Structural complexity does not automatically translate into hardship if courses are easily passed. We therefore complement graph metrics with **course-level empirical hardship measures** derived from student outcomes, conceptually related to blocking and delay factors in curricular analytics (Heileman et al., 2018).

For each course $j$ in a given degree–curriculum we compute:

1. **First-try pass probability** $p_j^{(1)}$.
   Proportion of students who pass $j$ on their first attempt.

2. **Eventual pass probability** $p_j^{(\infty)}$.
   Proportion of students who ever pass $j$ among those who attempt it at least once.

3. **Mean number of attempts** $\bar{a}_j$.
   Average count of attempts at $j$ among students who eventually pass.

4. **Dropout-after-fail rate** $d_j$.
   Proportion of students whose last recorded action in the degree–curriculum is a failure (or sequence of failures) in $j$ followed by permanent absence, adapted from progression analyses in curricular networks (Yang et al., 2025; Paz, 2025b).

5. **Blocking factor** $b_j$.
   For each successor course $k$ reachable from $j$ in the empirical DAG, we estimate the probability that a student attempts $k$ only after passing $j$. Summing across successors yields:

$$b_j = \sum_{k \in \text{succ}(j)} P(\text{attempt } k \mid \text{pass } j) \cdot \mathbb{1}\{j \text{ precedes } k\},$$

which approximates the number of downstream opportunities effectively gated by $j$. This mirrors blocking factors defined for formal curricular graphs (Krasnov, 2024).

We then define a **course-level BlockingScore** as:

$$\text{BlockingScore}_j = w_1 \left(1 - p_j^{(1)}\right) + w_2 \left(1 - p_j^{(\infty)}\right) + w_3 \bar{a}_j + w_4 d_j + w_5 b_j,$$

with non-negative weights $w_1, \ldots, w_5$ chosen so that each term contributes on a comparable scale (in practice we set all $w_k = 1$ after standardising the components). Higher values indicate courses that are difficult to pass, frequently failed, and structurally positioned to block access to many downstream courses.

### 3.4 Degree–curriculum hardship index

To characterise the hardness of an entire degree–curriculum, we aggregate both structural and empirical information. For each unit we compute:

1. **Mean BlockingScore** across all courses, and the proportion of courses in the top decile of BlockingScore.

2. **Empirical Hardship Index**

$$H_{\text{emp}} = z(\overline{\text{BlockingScore}}) + z(\%\text{ high-blocking courses}),$$

where bar denotes mean and $z$ indicates standardisation across all units.

3. **Composite Hardship Index**

$$H_{\text{comp}} = \frac{1}{2}(z(H_{\text{struct}}) + z(H_{\text{emp}})),$$

which gives equal weight to structural and empirical hardship.

This composite index serves as the main explanatory variable in subsequent analyses. We examine its relationship with degree–curriculum dropout rates and mean time to degree using descriptive comparisons and scatterplots rather than predictive modelling, in line with recent calls for structural, interpretable analytics that inform curriculum reform and policy rather than focusing solely on predictive accuracy (Qushem et al., 2025; Paz, 2025a; Rodríguez-Ortiz et al., 2025).

Together, these components define the "topology of hardship": a multicomponent representation of curriculum hardness grounded in empirical student flow and graph-theoretic structure.

## 4. RESULTS

### 4.1 Structural profile of empirical curriculum graphs

Across the 29 degree–curriculum units that passed the robustness filters described in Section 2.3, empirical curriculum graphs remain relatively sparse and shallow. Table 1 summarises the main structural metrics. On average, graphs contain about 45 subjects and 300 inferred dependencies, with wide dispersion in both size and connectivity. Densities cluster at low values (mean ≈ 0.13), and the median longest path is four steps, with only a few curricula exhibiting chains of eight or nine courses.

**Table 1.** Descriptive statistics for the 29 empirical curriculum graphs in the analytic sample. For each degree–curriculum unit, we report the number of subjects (nodes), number of inferred prerequisite edges, graph density and length of the longest prerequisite chain.

| Metric | count | mean | std | min | 25% | 50% | 75% | max |
| --- | --- | --- | --- | --- | --- | --- | --- | --- |
| n_subjects | 29.0 | 45.03 | 24.14 | 16 | 32 | 42 | 52 | 119 |
| n_edges | 29.0 | 289.34 | 314.27 | 4 | 20 | 176 | 398 | 1202 |
| density | 29.0 | 0.12 | 0.10 | 0.002 | 0.038 | 0.090 | 0.177 | 0.401 |
| longest_path | 29.0 | 4.03 | 1.80 | 2 | 2 | 4 | 5 | 9 |

Figure 1 provides a complementary view of the structural landscape. The left panel shows the distribution of the number of subjects, highlighting that most curricula fall between 30 and 60 courses, with a small subset of large, long-cycle programmes reaching more than 100 subjects. The middle panel displays graph density, which is heavily skewed towards low values: most curricula operate with densities below 0.20, confirming that students seldom face densely interlocked prerequisite structures. The right panel depicts the distribution of longest paths, indicating that while many curricula can be traversed in three to five steps, a non-trivial tail of plans still demands longer critical chains.

**Figure 1.** Distributions of structural metrics across the 29 empirical curriculum graphs: (a) number of subjects, (b) graph density and (c) length of the longest prerequisite chain.

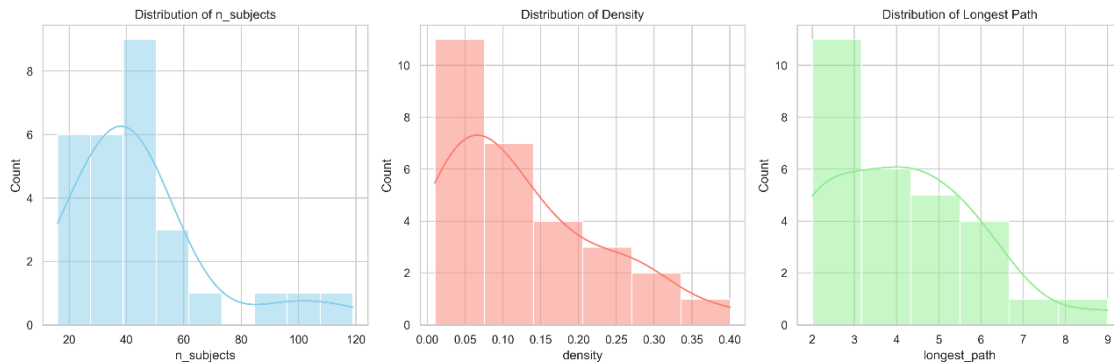

## 4.2 Bottleneck concentration and blocking patterns

Beyond global sparsity, several empirical graphs exhibit strong concentration of blocking around a small set of structurally central courses. Figure 2 illustrates this pattern for one representative case (Degree 54, Curriculum 1991). A single hub course generates outgoing edges towards a large portion of the curriculum, effectively acting as a structural "gatekeeper". Such configurations are precisely the type of topological feature that the blocking score is designed to capture.

**Figure 2.** Example empirical curriculum graph for Degree 54, Curriculum 1991. Arrow thickness reflects the frequency with which course A is observed before course B in students' trajectories, highlighting a highly central blocking hub.

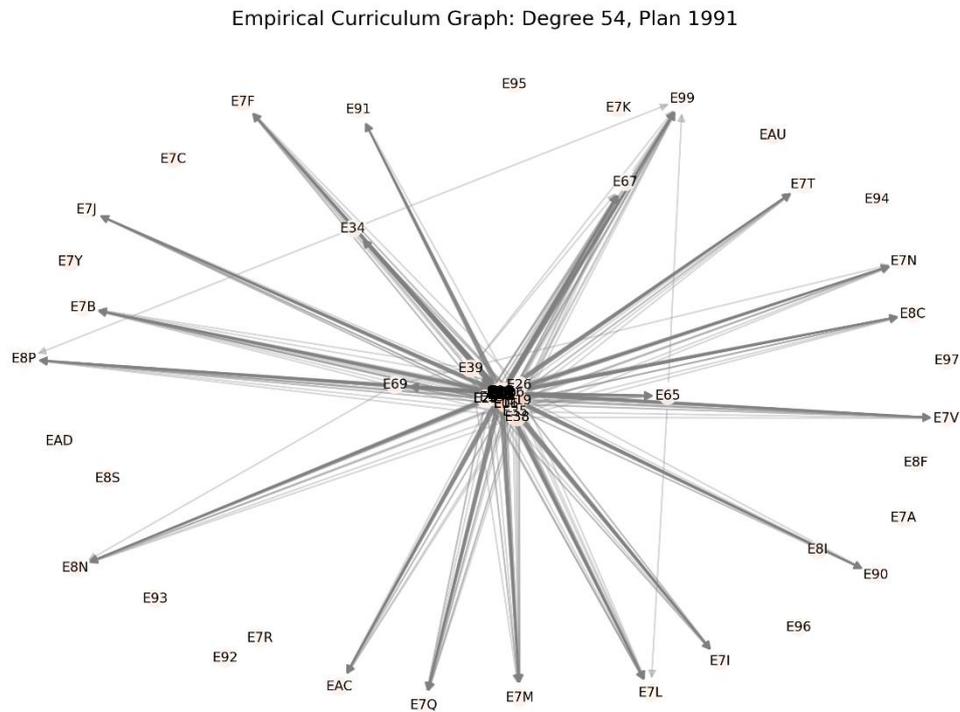

Table 2 ranks the degree–curriculum units in the analytic sample by composite hardship index. The highest values are concentrated in older and late-1990s plans: Ingeniería Civil (1996) and Programador Universitario (1996/2004) occupy the top positions (composite hardship up to ≈ 1.95), closely followed by Licenciatura en Matemática and Licenciatura en Física (both 1982). In all these cases, high composite hardship coincides with large mean BlockingScore values, often above 3.0, indicating that a small set of structurally central courses keeps a substantial fraction of students temporarily stalled.

**Table 2.** Degree–curriculum units in the analytic sample (N = 29), ranked by composite hardship index. For each unit, we report degree name, curriculum year, longest path length, mean blocking score and standardised hardship indices.

| degree_id | degree_name | curriculum_id | longest_path | mean_blocking_score | hardship_structural | hardship_empirical | hardship_composite |
|---|---|---|---|---|---|---|---|
| 47 | Ingeniería Civil | 1996 | 6 | 3.95 | 6 | 0.32 | 1.95 |
| 90 | Programador Universitario | 2004 | 5 | 1.70 | 5 | 0.39 | 1.93 |
| 82 | Licenciatura en Matemática | 1982 | 5 | 3.08 | 5 | 0.34 | 1.68 |
| 45 | Ingeniería Civil | 1996 | 9 | 4.02 | 9 | 0.17 | 1.57 |
| 90 | Programador Universitario | 1996 | 5 | 3.45 | 5 | 0.31 | 1.56 |
| 25 | Licenciatura en Física | 1982 | 2 | 1.37 | 2 | 0.77 | 1.54 |
| 55 | Ingeniería en Computación | 2004 | 6 | 1.33 | 6 | 0.25 | 1.47 |
| 51 | Ingeniería Eléctrica (CB) | 1991 | 2 | 0.90 | 2 | 0.68 | 1.36 |
| 33 | Tecnicatura Universitaria en Tecnología Azucarera e Industrias Derivadas | 1997 | 4 | 1.78 | 4 | 0.33 | 1.33 |
| 73 | Ingeniería Química | 2 | 4 | 0.67 | 4 | 0.32 | 1.28 |
| 55 | Ingeniería en Computación | 1991 | 6 | 2.64 | 6 | 0.21 | 1.27 |
| 52 | Ingeniería Electrónica | 1991 | 5 | 1.89 | 5 | 0.25 | 1.27 |
| 12 | Ingeniería en Agrimensura | 1998 | 4 | 2.20 | 4 | 0.29 | 1.15 |
| 98 | Diseñador de Iluminación | 2009 | 4 | 0.81 | 4 | 0.26 | 1.03 |
| 23 | Licenciatura en Física | 2001 | 4 | 2.18 | 4 | 0.25 | 1.02 |
| 97 | Técnico Diseñador Universitario en Iluminación | 2000 | 2 | 0.67 | 2 | 0.45 | 0.90 |
| 101 | Ingeniería Biomédica | 2002 | 3 | 1.88 | 3 | 0.29 | 0.86 |
| 97 | Técnico Diseñador | 1999 | 3 | 0.55 | 3 | 0.28 | 0.85 |

| | | | | | | |
|---|---|---|---|---|---|---|
| | Universitario en Iluminación | | | | | |
| 12 | Ingeniería en Agrimensura | 2007 | 3 | 0.40 | 3 | 0.27 | 0.81 |
| 16 | Ingeniería Geodésica y Geofísica | 2007 | 2 | 0.58 | 2 | 0.38 | 0.76 |
| 54 | Ingeniería Eléctrica | 1991 | 5 | 1.76 | 5 | 0.15 | 0.76 |
| 21 | Bachiller Universitario en Física | 2001 | 2 | 0.48 | 2 | 0.32 | 0.63 |
| 92 | Licenciatura en Informática | 2004 | 3 | 0.64 | 3 | 0.17 | 0.52 |
| 46 | Ingeniería Civil | 1996 | 4 | 1.40 | 4 | 0.12 | 0.49 |
| 26 | Licenciatura en Física | 1982 | 6 | 0.46 | 6 | 0.08 | 0.48 |
| 58 | Ingeniería en Computación | 2005 | 2 | 0.34 | 2 | 0.21 | 0.41 |
| 11 | Agrimensura | 1998 | 2 | 0.27 | 2 | 0.20 | 0.41 |
| 27 | Licenciatura en Física | 1982 | 7 | 0.48 | 7 | 0.06 | 0.39 |
| 24 | Tecnicatura Universitaria en Física Ambiental | 2011 | 2 | 0.16 | 2 | 0.09 | 0.19 |

Note. Degree names are kept in Spanish to reflect the official programme names used by the institution. The ranking is based on structural and empirical hardship metrics and is independent of the language of the labels.

The overall distribution of mean blocking scores is shown in Figure 3. Most curricula lie below 1.5, but a long right tail extends beyond 3.5. This tail is populated precisely by the historical plans identified in Table 2, reinforcing the interpretation of these designs as structurally "heavy" curricula where a few courses exert disproportionate blocking power.

**Figure 3.** Distribution of mean blocking scores across the 29 empirical curriculum graphs. Higher values indicate a larger fraction of students temporarily stalled by structurally central courses.

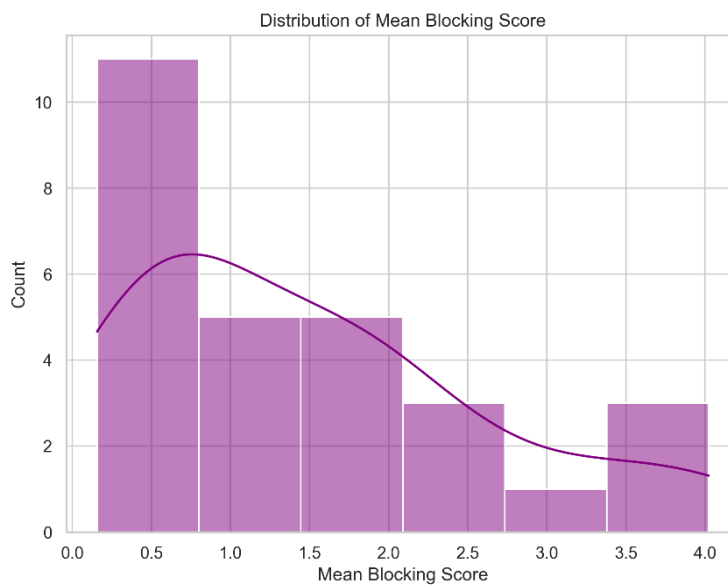

### 4.3 Distribution of hardship indices

Table 3 reports summary statistics for the three hardship indices—structural, empirical and composite—across the 29 curricula. As expected for standardised indices, means are essentially zero and standard deviations are close to one. Structural hardship spans from approximately −1.13 to +2.76, empirical hardship from −1.47 to +3.14, and composite hardship from −1.73 to +1.89. The broader spread of empirical hardship reflects the heterogeneity of observed failure and repetition patterns even under similar topologies, while the composite index blends both dimensions into a single, moderately dispersed scale.

**Table 3.** Descriptive statistics for standardised hardship indices (z-scores) across the 29 degree–curriculum units in the analytic sample.

|  | count | mean | std | min | 25% | 50% | 75% | max |
|---|---|---|---|---|---|---|---|---|
| hardship_structural | 29.0 | 0.0 | 1.0 | -1.129 | -1.129 | -0.019 | 0.536 | 2.755 |
| hardship_empirical | 29.0 | -0.0 | 1.0 | -1.469 | -0.522 | -0.092 | 0.268 | 3.135 |
| hardship_composite | 29.0 | -0.0 | 1.0 | -1.729 | -0.818 | -0.02 | 0.687 | 1.885 |

Taken together, Figures 2 and 3 and Tables 2 and 3 show that hardship is not evenly distributed: a minority of historical curricula concentrate both high structural blocking and unfavourable empirical performance, producing composite hardship levels two standard deviations above the institutional mean.

### 4.4 Hardship and institutional outcomes

Finally, we examine whether harder curricula—in the sense of higher composite hardship—are associated with poorer institutional outcomes. Figure 4 presents scatterplots of composite hardship against dropout rate and mean time to degree for the 29 degree–curriculum units in the analytic sample. A simple linear fit suggests a modest positive association between composite hardship and dropout rate (Pearson r ≈ 0.23): curricula with higher hardship tend, on average, to exhibit higher proportions of students leaving without a degree. The dispersion is substantial, but the trend is consistent with the notion of structural hardness as a "dropout amplifier".

By contrast, the right-hand panel of Figure 4 shows no clear linear relationship between composite hardship and mean time to degree (r ≈ −0.06). Confidence bands around the fitted line are wide and centred around a nearly flat slope, indicating that harder curricula do not systematically prolong the trajectories of those who eventually graduate. Instead, they appear more likely to push students out of the system altogether, while survivors progress at similar speeds to their peers in less demanding topologies.

**Figure 4.** Relationship between composite hardship and institutional outcomes across the 29 degree–curriculum units. Left: dropout rate. Right: mean time to degree (years). Each point represents one degree–curriculum unit; lines show least-squares fits with 95% confidence bands (N = 29 in both panels).

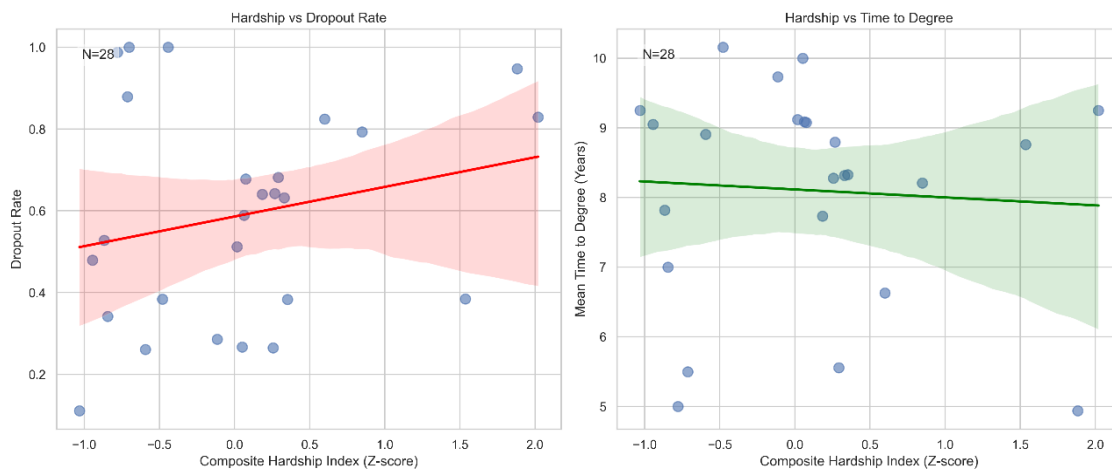

## 5. DISCUSSION

### 5.1 Curriculum hardness as a structural property

The results show that "hard" programmes are not just those with difficult exams or demanding lecturers. Hardness emerges as a *topological property* of degree–curriculum graphs: dense prerequisite structure, long critical chains, and high bottleneck concentration push the Structural Hardship Index upwards, while repeated failures and high blocking scores drive the Empirical Hardship Index. Together they produce a Composite Hardship Index that aligns with dropout and with wider dispersion in time to degree.

This connects directly with the idea of curricular complexity in the curricular analytics framework, where structural features such as delay and blocking factors are linked to student success (Heileman et al., 2018). Our results can be read as an empirical, student-flow-based instantiation of that framework: rather than starting from formal prerequisite rules, we reconstruct *effective* dependencies from longitudinal trajectories and then layer empirical difficulty on top of structural complexity.

From a network perspective, the findings are consistent with prior work showing that some curricula are shallow and loosely coupled, whereas others are deep and tightly interdependent (Simon de Blas et al., 2021; Zuev, 2025). The long tail of high-hardship degree–curricula we observed corresponds precisely to those deep and dense configurations, now shown to be associated with more volatile retention outcomes.

**5.2 Hardship as a structural amplifier of risk**

The relationship between the Composite Hardship Index and institutional outcomes is not perfectly linear, but the pattern is clear: high-hardship degree–curricula tend to show higher dropout and more dispersion in completion times. This supports interpreting hardness as a *structural amplifier of risk*: it does not "cause" dropout on its own, but it magnifies the consequences of any weakness in preparation, support, or macro-context.

Within the broader CAPIRE framework, the topology of hardship sits naturally at the N2/N3 interface, where academic history and curricular friction are modelled as part of a leakage-aware trajectory layer (Paz, 2025b). Earlier CAPIRE work emphasised how structural friction and macro-shocks (strikes, inflation) interact with student trajectories; the present analysis shows that even when we hold the macro-context fixed, some degree–curricula are structurally more exposed than others.

The strong concentration of BlockingScore in a small subset of "hard blockers" is particularly relevant. It mirrors previous findings that a handful of high-failure courses can shape entire dropout pathways (Heileman et al., 2018), but here we add a structural nuance: these courses are not only difficult, they also sit in positions where failure closes many downstream options. In other words, they have both *high local difficulty* and *high global leverage*.

**5.3 Implications for curriculum design and accreditation**

For curriculum design, the message is uncomfortable but operational: some degree–curricula are objectively more constrained and risk-amplifying than others, and this can be quantified. Network-based curriculum studies have already argued that graph metrics can inform the redesign of study plans (Simon de Blas et al., 2021; Zuev, 2025). Our contribution is to show how those structural metrics can be combined with empirical hardship to rank and compare degree–curricula within a single institution.

In practical terms, accreditation processes and internal quality assurance could use the Composite Hardship Index and its components as diagnostic indicators:

- Identify degree–curricula with extreme structural hardship and review whether long critical chains or dense blocks are pedagogically justified.

- Flag courses with simultaneously high BlockingScore and high betweenness as candidates for redesign, support interventions, or redistribution of content.

- Compare hardship profiles across cohorts and plan versions to document whether reforms actually reduce structural risk, rather than just shifting it.

This aligns with calls in curricular analytics to move from anecdotal perceptions of "hard" programmes to measurable, reproducible indicators that can be monitored over time (Heileman et al., 2018). The fact that our graphs are derived from empirical trajectories rather than catalogue descriptions also makes the indicators more defensible in internal negotiations: they reflect what students actually do, not what the handbook imagines they will do.

**5.4 Positioning within the learning analytics landscape**

Most current learning analytics work in higher education focuses on predictive models of performance, engagement or dropout at course or programme level, often using machine learning and, increasingly, generative AI (Qushem et al., 2025; Rodríguez-Ortiz et al., 2025). These models are useful for early warning and personalised feedback, but they typically treat programme structure as a background condition rather than as an object of analysis.

The topology of hardship flips that perspective: instead of asking "which students are at risk in this programme?", we ask "which programmes are structurally more likely to put students at risk, and where exactly does that risk concentrate?". The result is complementary to predictive risk scores:

- Predictive models can tell us *who* is likely to struggle and *when*.

- Topology-of-hardship metrics tell us *where* the structure itself amplifies struggle and *how* it does so.

Within a leakage-aware, multilevel architecture such as CAPIRE (Paz, 2025a), the two layers can be combined cleanly: structural hardness becomes an N2/N3 feature that shapes risk, while N1 socio-economic variables and N4 macro-context capture upstream and downstream constraints. This responds directly to recent reviews that call for LA systems which are not only accurate but also theory-informed, transparent and usable for institutional decision-making (Qushem et al., 2025; Rodríguez-Ortiz et al., 2025).

The practical implication is simple: any institution with longitudinal registration data can reconstruct empirical curriculum graphs, compute hardship indices, and use them to prioritise structural reforms. This is low-hanging fruit compared with deploying full causal models or agent-based simulations, and yet it already shifts the focus from blaming students to interrogating the "shape of the game board" itself.

## 6. CONCLUSIONS

### 6.1 Summary of findings

This study introduced the notion of the **topology of hardship** as a way to treat curriculum hardness as a measurable, structural property of engineering degrees. Using longitudinal registry data from a large public university, we reconstructed empirical curriculum graphs for a robust analytic sample of 29 degree–curriculum units with complete outcome data, derived hardship metrics, and related the topology of hardship to dropout and time-to-degree, not from handbooks or formal prerequisite tables, but from actual student trajectories. On these graphs we computed structural metrics (size, density, longest path, bottleneck concentration) and empirical hardship metrics (BlockingScore, blocking prevalence), and combined them into a **Composite Hardship Index**.

The results show that engineering curricula are far from homogeneous. Most degree–curricula exhibit shallow, sparsely connected structures with low hardship, while a smaller subset displays dense dependency patterns, long critical chains and high bottleneck concentration. In these structurally demanding programmes, empirical hardship is also higher: failures cluster in a small set of high-leverage courses, mean BlockingScore rises, and students' progression becomes more fragile. When aggregated at degree–curriculum level, the Composite Hardship Index aligns with higher dropout rates and greater dispersion in time to degree, suggesting that curriculum hardness behaves as a **structural amplifier of risk** rather than as a purely symbolic label.

## 6.2 Implications for practice and policy

Methodologically, the paper extends work in curricular analytics and course-prerequisite networks by grounding curriculum graphs in **empirical flow** instead of declared prerequisites (Heileman et al., 2018; Simon de Blas et al., 2021). This matters for institutions where formal rules are loosely enforced or frequently negotiated. The topology of hardship provides a bridge between network-level measures and institutional outcomes: it offers a compact way of ranking and comparing degree–curricula and of pinpointing where structural pressure concentrates.

Practically, the framework can be used to inform curriculum redesign, accreditation and internal quality assurance. Degree–curricula with extreme hardship values can be flagged for review; high-blocking, high-betweenness courses can be prioritised for pedagogical support, assessment redesign or redistribution of content; and successive plan reforms can be evaluated in terms of their impact on structural and empirical hardship rather than only on descriptive statistics such as credit load or nominal duration. Within a broader learning analytics ecosystem, hardship metrics complement individual-level predictive models by shifting the focus from "which students are at risk?" to "which programmes and courses systematically amplify risk, and through which structural mechanisms?" (Qushem et al., 2025; Rodríguez-Ortiz et al., 2025).

For frameworks such as CAPIRE, the contribution is conceptual as well as technical. The topology of hardship operationalises the N2 layer—curriculum and institutional structure—as more than a fixed background: it becomes a configurable, quantifiable object that can be linked to survival analysis, macro-shock modelling and intervention simulation (Paz, 2025b). This opens the door to policy scenarios where structural changes to the degree plan are treated explicitly as levers in retention engineering, rather than as cosmetic adjustments.

## 6.3 Limitations and directions for further research

The analysis has several limitations that suggest directions for future work. First, it is based on **a single institution** and a specific set of engineering degrees; the distribution of hardness and its relationship with dropout may look different in shorter programmes, professionally oriented curricula or institutions with strict prerequisite enforcement. Secondly, the reconstruction of empirical curriculum graphs depends on threshold choices for edge inference and on minimum size criteria for accepting a graph as non-degenerate. Alternative thresholds, or Bayesian treatments of edge uncertainty, could be explored to assess robustness.

Thirdly, the topology of hardship is **observational and descriptive**: it captures how structures and difficulties co-occur, but cannot, on its own, separate the effects of curriculum design from those of pedagogy, assessment cultures or student support. Integrating the hardship indices into explicit causal models or agent-based simulations—as envisaged in the wider CAPIRE research agenda—would allow more precise estimation of how structural changes might affect trajectories under different policy assumptions. Finally, while this study focused on composite hardship, future work could examine specific archetypes of hard curricula (e.g., long-chain versus hub-and-spoke structures) and investigate whether different structural "modes of hardness" interact differently with students' backgrounds and learning environments.

Despite these limitations, the core message is robust: **curriculum hardness is measurable**, it varies substantially across degree–curricula within the same institution, and it aligns with real differences in how cohorts move, stall and leave. Treating the topology of hardship as a first-class analytical object shifts the conversation from individual deficits to structural responsibility and provides a concrete starting point for data-informed curriculum reform.